\documentclass[showpacs,preprintnumbers,amsmath,amssymb,preprint]{revtex4}

\usepackage{graphicx}

\usepackage{dcolumn}

\usepackage{bm}

\begin{document}

\title{Relativistic Compact Objects in Isotropic Coordinates}

\author{M. K. Mak}

\email{mkmak@vtc.edu.hk}\affiliation{Department of Physics, The
University of Hong Kong, Pok Fu Lam Road, Hong Kong SAR, P. R.
China}

\author{T. Harko}
\email{harko@hkucc.hku.hk}\affiliation{Department of Physics, The
University of Hong Kong, Pok Fu Lam Road, Hong Kong SAR, P. R.
China}

\date{\today}

\begin{abstract}
We present a matrix method for obtaining new classes of exact
solutions for Einstein's equations representing static perfect
fluid spheres. By means of a matrix transformation, we reduce
Einstein's equations to two independent Riccati type differential
equations for which three classes of solutions are obtained. One
class of the solutions corresponding to the linear barotropic type
fluid with an equation of state $p=\gamma \rho $ is discussed in
detail.
\end{abstract}



\maketitle

\section{Introduction}

Relativistic stellar models have been studied ever since the first
solution of Einstein's field equation was obtained by
Schwarzschild for the interior of a compact object in hydrostatic
equilibrium. The search for the exact solutions is continuously of
an interest to physicists. The most general exact solution for
Vaidya-Tikekar isentropic superdense star was obtained by Gupta
and Jasmin  \cite{GuJa00}.  Hernandez and Nunez \cite{HeNu04}
presented a general method for obtaining static anisotropic
spherically symmetric solutions satisfying a nonlocal equation of
state from known density profiles. Several new exact solutions for
anisotropic stars of constant density were presented in Dev and
Gleiser \cite{DeGl02}. In \cite{Fo00} an algorithm was proposed to
generate any number of physically realistic pressure and density
profiles for spherical perfect fluid distributions without
evaluating integrals. The gravitational field equations for static
stellar models with a linear barotropic equation of state and with
a polytropic equation of state $p=k\rho ^{1+1/n}$ were recast
respectively into two complementary 3-dimensional regular system
of ordinary differential equations on compact state space
\cite{NiUg01a}, \cite{NiUg01b}. These systems were analyzed
numerically and qualitatively, using the theory of dynamical
systems. Schmidt and Homann \cite{ScHo00} discussed numerical
solutions of Einstein's field equation describing static
spherically symmetric conglomerations of a photon star with an
equation of state $\rho =3p$. Upper limits for the mass-radius
ratio were derived for compact general relativistic objects in the
presence of a cosmological constant in \cite{MaDo00} and for the
charged general relativistic fluid spheres in \cite{MaHaDo01}. The
effect of a cosmological constant on the structure of the general
relativistic objects was analyzed by B\"ohmer \cite{Bo04}. The
equations describing the adiabatic, small radial oscillations of
general relativistic stars have been generalized to include the
effects of a cosmological constant in \cite{BoHa05}. Interior
solutions for charged and neutral anisotropic fluid spheres,
satisfying all the required physical conditions, were obtained in
\cite{HaMa00} and \cite{MaHa03}. Excellent reviews of the
relativistic static fluid spheres were given in Finch and Skea
\cite{FiSk95}, Herrera and Santos \cite{HeSa97} and Delgaty and
Lake \cite{DeLa98}.

For a static fluid configuration Einstein's equations represent an
under-determined system of nonlinear ordinary differential
equations of the second order. For the special case of a static
isotropic perfect fluid, the field equations of Einstein's theory
can be reduced to a set of three coupled ordinary differential
equations in four unknowns. To obtain a realistic stellar model,
one can start with an equation of state. Such input equations of
state do not normally allow for closed form solutions. In arriving
to exact solutions, one can solve the field equations by making an
ad hoc assumption for one of the metric functions or the energy
density, hence the equation of state being computed from the
resulting metric.

It is the purpose of the present paper to present a matrix method
for obtaining new classes of exact solutions for Einstein's
equations representing static perfect fluid spheres. By means of a
matrix transformation, we reduce Einstein's equations to two
independent Riccati's differential equations for which three
classes of solutions are obtained. One class of the solutions
corresponding to the linear barotropic type fluid with an equation
of state $p=\gamma \rho $ is discussed in detail.

The present paper is organized as follows. In Section II we reduce
the basic equation describing the interior stellar evolution to a
matrix equation which is separated into two independent Riccati's
differential equations. Three classes of solutions of the
gravitational field equations are presented and the stiff solution
is discussed in Section III. We conclude our results in Section
IV.

\section{Field equations and the matrix method}

For a relativistic fluid sphere model the line element inside the
matter distribution takes the form
\begin{equation}\label{1}
ds^{2}=e^{\nu \left( r\right) }dt^{2}-e^{\lambda \left( r\right)
}\left( dr^{2}+r^{2}d\Omega ^{2}\right) ,
\end{equation}
where $d\Omega ^{2}=d\theta ^{2}+\sin ^{2}\theta d\phi ^{2}.$

Assuming that the matter content inside the spheres is a perfect
fluid, the field equations take the form \cite{LaLi}
\begin{equation}\label{2}
8\pi p=e^{-\lambda }\left[ \frac{\left( \lambda ^{^{\prime }}\right) ^{2}}{4}%
+\frac{\lambda ^{^{\prime }}\nu ^{^{\prime }}}{2}+\frac{\lambda
^{^{\prime }}+\nu ^{^{\prime }}}{r}\right] ,
\end{equation}
\begin{equation}\label{3}
8\pi p=e^{-\lambda }\left[ \frac{\lambda ^{^{^{\prime \prime }}}}{2}+\frac{%
\nu ^{^{\prime \prime }}}{2}+\frac{\left( \nu ^{^{\prime }}\right) ^{2}}{4}+%
\frac{\lambda ^{^{\prime }}+\nu ^{^{\prime }}}{2r}\right] ,
\end{equation}
\begin{equation}\label{4}
8\pi \rho =-e^{-\lambda }\left[ \lambda ^{^{^{\prime \prime
}}}+\frac{\left( \lambda ^{^{\prime }}\right)
^{2}}{4}+\frac{2\lambda ^{^{\prime }}}{r}\right] ,
\end{equation}
where a prime denotes $\frac{d}{dr}$.

By introducing the transformation \cite{HaHa94}
\begin{equation}\label{5}
R=\ln r,
\end{equation}
equations (\ref{2})-(\ref{4}) become
\begin{equation}\label{6}
8\pi p=\frac{e^{-\lambda }}{r^{2}}\left[ \frac{\left(
\dot{\lambda}\right)
^{2}}{4}+\frac{\dot{\lambda}\dot{\nu}}{2}+\dot{\lambda}+\dot{\nu}\right]
,
\end{equation}
\begin{equation}\label{7}
8\pi p=\frac{e^{-\lambda }}{r^{2}}\left[ \frac{\ddot{\lambda}+\ddot{\nu}}{2}+%
\frac{\left( \dot{\nu}\right) ^{2}}{4}\right] ,
\end{equation}
\begin{equation}\label{8}
8\pi \rho =-\frac{e^{-\lambda }}{r^{2}}\left[
\ddot{\lambda}+\frac{\left( \dot{\lambda}\right)
^{2}}{4}+\dot{\lambda}\right] ,
\end{equation}
where the dot denotes $\frac{d}{dR}$. The set of coordinates
resulting after the re-scaling of the radial coordinate $r$ by
means of the transformation (\ref{5}) are called isotropic
coordinates.

From the isotropic pressure condition, we obtain the following
basic equation describing the structure of general relativistic
stellar type objects in isotropic coordinates:
\begin{equation}\label{9}
\ddot{\lambda}+\ddot{\nu}+\frac{1}{2}\left[ \left(
\dot{\nu}\right) ^{2}-\left( \dot{\lambda}\right) ^{2}\right]
-\dot{\lambda}\dot{\nu}-2\left( \dot{\nu}+\dot{\lambda}\right) =0.
\end{equation}

By means of the following transformations
\begin{equation}\label{10}
\lambda =\int udR,\nu =\int wdR,
\end{equation}

and with the use of the two matrices
\begin{equation}\label{11}
L=\left[ \left( 2\dot{u}-4u\right) +\left( 2\dot{w}-4w\right)
\right] ,
\end{equation}
Equation (\ref{9}) can be written as a matrix equation in the form
\begin{equation}\label{12}
L=M^{T}AM.
\end{equation}

We introduce now two new variables $u_{-}$ and $u_{+}$ which are
obtained by means of a linear transformation, described by the
matrix $K$ and applied to the matrix $N$,
\begin{equation}\label{13}
M=KN=\frac{1}{\sqrt{2}}\left(
\begin{array}{cc}
\left( 1-\sqrt{2}\right) m_{-}^{-1} & \left( 1+\sqrt{2}\right) m_{+}^{-1} \\
m_{-}^{-1} & m_{+}^{-1}%
\end{array}
\right) \left(
\begin{array}{c}
u_{-} \\
u_{+}%
\end{array}
\right) .
\end{equation}

In the new variables, equation (\ref{12}) reduces to $\ $the form
$L=N^{T}K^{T}AKN$. We choose the elements of the matrix $K$ so
that

\[
K^{T}AK=\left(
\begin{array}{cc}
-\sqrt{2} & 0 \\
0 & \sqrt{2}%
\end{array}
\right) .
\]
As a result of these transformations the right hand side of the
matrix equation (\ref{12}) can be diagonalized, and the resulting
equations satisfied by the new unknown functions $u_{-}$, $u_{+}$
are
\begin{equation}\label{14}
m_{-}\frac{du_{-}}{dR}-2m_{-}u_{-}+u_{-}^{2}=-m_{+}\frac{du_{+}}{dR}%
+2m_{+}u_{+}+u_{+}^{2}=F\left( R\right) ,
\end{equation}
where $m_{\mp }=\sqrt{2\mp \sqrt{2}}$ and a solution generating function $%
F\left( R\right) $ is introduced.

\section{Three classes of solutions of the static field equations}

In the previous Section, by means of a matrix transformation, the
cross-term $\lambda \nu $ cancelled from the isotropic pressure
equation, written in the terms of new variables $u_{-}$ and
$u_{+}$ Hence, we have reduced the second order differential
equation (\ref{9}) to two independent Riccati type differential
equations, for which exact solutions of the field equations
describing the stellar structure can be generated by the
appropriate choices of the generating function $F\left( R\right)
$.

\subsection{Case A:}
\[
F(R) =\beta ={\rm constant}
\]

In order to obtain some solutions of the gravitational field
equations for our stellar model, we assume the generating function
$F\left( R\right) =\beta =$constant. In this case equations
(\ref{14}) take the form

\begin{equation}\label{15}
m_{-}\frac{du_{-}}{dR}-2m_{-}u_{-}+u_{-}^{2}=\beta ,-m_{+}\frac{du_{+}}{dR}%
+2m_{+}u_{+}+u_{+}^{2}=\beta ,
\end{equation}

Depending on the sign of the parameter $\beta $, we obtain two
distinct classes of solutions of Eq. (\ref{15})

(i) $\Delta _{\mp }>0$

\begin{equation}\label{16}
u_{\mp }=\frac{\sqrt{\Delta _{\mp }}+2\varepsilon _{\pm }m_{\mp
}+\left(
\sqrt{\Delta _{\mp }}-2\varepsilon _{\pm }m_{\mp }\right) \exp \left[ \frac{%
\sqrt{\Delta _{\mp }}}{\varepsilon _{\pm }m_{\mp }}\left( R_{\mp }-R\right) %
\right] }{2\left[ 1-\exp \left[ \frac{\sqrt{\Delta _{\mp
}}}{\varepsilon _{\pm }m_{\mp }}\left( R_{\mp }-R\right) \right]
\right] },
\end{equation}

(ii) $\Delta _{\mp }<0$

\begin{equation}\label{17}
u_{\mp }=\varepsilon _{\pm }m_{\mp }+\frac{\sqrt{-\Delta _{\mp }}}{2}\tan %
\left[ \frac{\sqrt{-\Delta _{\mp }}}{2\varepsilon _{\pm }m_{\mp
}}\left( R_{\mp }-R\right) \right] ,
\end{equation}
where $\Delta _{\mp }=4\left( m_{\mp }^{2}+\beta \right) $,
$\varepsilon _{\mp }=\mp 1$, $R_{\pm }$ are arbitrary constants of
integration.

\subsection{Case B:}
\[
F\left( R\right) =-2m_{-}u_{0-}+u_{0-}^{2}=2m_{+}u_{0+}+u_{0+}^{2}
\]

Assuming that the solutions of equations (\ref{14}) are constants,
i. e., $u_{\mp }=u_{0\mp }=$constant, we can obtain another
solution of the field equations for the static stellar model with
a linear barotropic equation of state.

With the use of Eq. (\ref{11}), equations (\ref{10}) become
\begin{equation}\label{18}
\lambda =\frac{1}{\sqrt{2}}\int \left[ \left( 1-\sqrt{2}\right) \frac{u_{-}}{%
m_{\_}}+\left( 1+\sqrt{2}\right) \frac{u_{+}}{m_{+}}\right] dR,
\end{equation}
\begin{equation}\label{19}
\nu =\frac{1}{\sqrt{2}}\int \left( \frac{u_{-}}{m_{\_}}+\frac{u_{+}}{m_{+}}%
\right) dR.
\end{equation}

The explicit forms of the metric functions $e^{\lambda }$ and
$e^{\nu }$ representing the relativistic fluid spheres for three
distinct classes of solutions are given by

$Case$ $(A)$ (i)
\begin{equation}\label{20}
e^{\lambda }=\lambda _{0}r^{\frac{1-\sqrt{2}}{2\sqrt{2}}\left( \frac{\sqrt{%
\Delta _{-}}}{m_{-}}+2\right) +\frac{1+\sqrt{2}}{2\sqrt{2}}\left( \frac{%
\sqrt{\Delta _{+}}}{m_{+}}-2\right) }\left[ e^{\frac{\sqrt{\Delta _{-}}}{%
m_{-}}\left( R_{-}-\ln r\right) }-1\right] ^{\frac{1-\sqrt{2}}{\sqrt{2}}}%
\left[ e^{-\frac{\sqrt{\Delta _{+}}}{m_{+}}\left( R_{+}-\ln r\right) }-1%
\right] ^{-\frac{1+\sqrt{2}}{\sqrt{2}}},
\end{equation}
\begin{equation}\label{21}
e^{\nu }=\nu _{0}r^{\frac{1}{2\sqrt{2}}\left( \frac{\sqrt{\Delta _{+}}}{m_{+}%
}+\frac{\sqrt{\Delta _{-}}}{m_{-}}\right) }\left[ e^{\frac{\sqrt{\Delta _{-}}%
}{m_{-}}\left( R_{-}-\ln r\right) }-1\right] ^{\frac{1}{\sqrt{2}}}\left[ e^{-%
\frac{\sqrt{\Delta _{+}}}{m_{+}}\left( R_{+}-\ln r\right) }-1\right] ^{-%
\frac{1}{\sqrt{2}}},
\end{equation}

$Case$ $(A)$\ (ii)

\begin{equation}\label{22}
e^{\lambda }=\lambda _{0}r^{-2}\cos
^{\frac{1-\sqrt{2}}{\sqrt{2}}}\left[
\frac{\sqrt{-\Delta _{-}}}{2m_{-}}\left( R_{-}-\ln r\right) \right] \cos ^{-%
\frac{1-\sqrt{2}}{\sqrt{2}}}\left[ -\frac{\sqrt{-\Delta
_{+}}}{2m_{+}}\left( R_{+}-\ln r\right) \right] ,
\end{equation}

\begin{equation}\label{23}
e^{\nu }=\nu _{0}\cos ^{\frac{1}{\sqrt{2}}}\left[ \frac{\sqrt{-\Delta _{-}}}{%
2m_{-}}\left( R_{-}-\ln r\right) \right] \cos ^{-\frac{1}{\sqrt{2}}}\left[ -%
\frac{\sqrt{-\Delta _{+}}}{2m_{+}}\left( R_{+}-\ln r\right)
\right] ,
\end{equation}

$Case$ $(B)$

\begin{equation}\label{24}
e^{\lambda }=\lambda _{0}r^{\frac{1}{\sqrt{2}}\left[ \left( 1-\sqrt{2}%
\right) a+\left( 1+\sqrt{2}\right) b\right] },
\end{equation}

\begin{equation}\label{25}
e^{\nu }=\nu _{0}r^{\frac{1}{\sqrt{2}}\left( a+b\right) },
\end{equation}
where we have denoted $a=u_{0-}m_{-}^{-1}$ and
$b=u_{0+}m_{+}^{-1}$.

For $Case$ $(B)$, the physical quantities, the energy density
$\rho $ and the fluid pressure $p$ are expressed in the following
form

\begin{eqnarray}\label{26}
8\pi \rho &=&\lambda _{0}^{-1}\left\{ \left( \frac{5}{8}-\frac{1}{\sqrt{2}}%
\right) a^{2}+\left( -1+\sqrt{2}+\frac{b}{4}\right) a+\left[ 1+\sqrt{2}%
+\left( \frac{5}{8}+\frac{1}{\sqrt{2}}\right) b\right] b\right\} \times \\
&&r^{-\left\{ \frac{1}{\sqrt{2}}\left[ \left( 1-\sqrt{2}\right) a+\left( 1+%
\sqrt{2}\right) b\right] \right\} -2},  \nonumber
\end{eqnarray}

\begin{equation}\label{27}
p=\frac{-8a+5a^{2}+8b+2ab+5b^{2}-4\sqrt{2}\left( a-b-2\right)
\left( a+b\right) }{8a-3a^{2}-8b+2ab-3b^{2}+2\sqrt{2}\left(
a-b-2\right) \left( a+b\right) }\rho =\gamma \rho ,
\end{equation}
where $0\leq \gamma \leq 1$.

Therefore we have obtained an exact solution of the field
equations for the perfect fluid sphere in isotropic coordinate
with a linear barotropic type equation of state.

By choosing the parameters $u_{0+}=0,u_{0-}=2m_{-}$ or
$u_{0-}=0,u_{0+}=-2m_{+}$, the metric functions given by Eqs.
(\ref{24}) and (\ref{25}) generates the line element

\begin{equation}\label{28}
ds^{2}=r^{\pm \sqrt{2}}dt^{2}-r^{\pm \sqrt{2}-2}\left(
dr^{2}+r^{2}d\Omega ^{2}\right) ,
\end{equation}
with the equation of state $p=\rho $ (stiff or Zeldovich equation
of state) for the dense matter.

It can be shown that the + and - signs are equivalent through the
transformation $r\longrightarrow r^{-1}.$ Therefore by using the
matrix method we have rediscovered this exact solution for the
stiff fluid matter that was obtained by Haggag and Hajj-Boutros
\cite{HaHa94}.

To overcome the difficulty of infinite energy density at the
center of the sphere, it is assumed that the matter distribution
has a core of radius $r_{0}$ and constant density $\rho _{0}$
which is surrounded by the fluid with pressure equal to energy
density. In view of
equations (\ref{24}) and (\ref{25}), by choosing $b=a$ we obtain $e^{\lambda }\sim e^{\nu }r^{\frac{1}{%
\sqrt{2}}\left( b-a\right) }$, or equivalently, $u_{0\mp
}=-2\sqrt{4\mp 2\sqrt{2}}$. Therefore the interior metric is
conformally flat satisfying the zero condition of the Weyl tensor.

By means of the transformation
$r^{\frac{a}{\sqrt{2}}+1}\longrightarrow r$, the conformally flat
metric becomes
\begin{equation}\label{29}
ds^{2}=r^{\frac{2a}{a+\sqrt{2}}}dt^{2}-\frac{2}{\left( a+\sqrt{2}\right) ^{2}%
}dr^{2}-r^{2}d\Omega ^{2}.
\end{equation}

In Schwarzschild coordinates Tolman \cite{To39} presented this
solution over sixty years ago. However, Gurses and Gursey
\cite{GuGu75} proved that the Schwarzschild interior metric for a
non-charged sphere is the {\it unique} solution for the Einstein's
gravitational field equations which is static and conformally
flat.

Surprisingly, the line element (\ref{29}) contradicts the results
suggested by Gurses and Gursey \cite{GuGu75}.

\section{Discussions and final remarks}

In the present paper we have studied general relativistic stellar
models described by the line element (\ref{1}), and have presented
two new classes of solutions, given by Eq. (\ref{20})-(\ref{23}),
of the field equations representing the interior structure of the
compact objects in the framework of general relativity. The class
of the solution corresponding to the stiff fluid has been
discussed in detail. From the physical point of view, the
mathematical solutions must satisfy certain physical requirements
to render them physically meaningful. The following conditions or
requirements have been generally accepted \cite{HeSa97}.

(a) the energy density $\rho $ and the pressure $p$ should be
positive and finite;

(b) their gradients $\frac{d\rho }{dr}$ and $\frac{dp}{dr}$ should
be negative;

(c) the speed of sound should be less than the speed of light.

(d) a physically reasonable energy-momentum tensor must obey a
trace condition, $\rho \geq 3p$.

(e) the interior metric should be joined continuously with the
exterior Schwarzschild metric.

(f) the pressure $p$ should vanish at the vacuum boundary of the
sphere.

(g) the structures should be stable under radial pulsations.

Not many exact solutions for relativistic fluid spheres are known
in isotropic coordinates. Hence, the main purpose of this paper is
to present the matrix method for obtaining the exact solutions of
the field equations for the compact objects in isotropic
coordinates. Eq. (\ref{14}) is a first order differential equation in two unknowns $%
u_{-}$and $u_{+}$. To obtain its solutions a solution generating
function $F\left( R\right) $  has been introduced in Eq.
(\ref{14}). Because of the mathematical structure of the master
equation (\ref{14}), hosts of solutions can be found by the
appropriate choice of the generating function $F\left( R\right) $
or by
making an ad hoc assumption for one of the variables $u_{-}$or $%
u_{+} $.

In order to have physically realistic stellar models, one needs to
test the solutions obtained via this method to ensure that they
 satisfy all the physical conditions (a)-(g).

\label{lastpage}


\begin{thebibliography}{99}


\bibitem{GuJa00} Gupta Y. K., Jasim M. K., 2000, Astrophys. Space Science, 272, 403
(2000).

\bibitem{HeNu04} Hernandez, H., Nunez, L. A., 2004, Can. J. Phys., 82,
29

\bibitem{DeGl02} Dev, K., Gleiser, M., 2002, Gen. Rel. Grav., 34, 1793

\bibitem{Fo00} Fodor, G., 2000, preprint gr-qc/0011040

\bibitem{NiUg01a} Nilsson, U. F., Uggla, C., 2001a, Annals. Phys., 286, 278

\bibitem{NiUg01b} Nilsson, U. F., Uggla, C., 2001b, Annals. Phys., 286,
292

\bibitem{ScHo00} Schmidt, H. J., Homann, F., 2000, Gen. Rel. Grav., 32, 919

\bibitem{MaDo00} Mak, M. K., Dobson, P. N., Jr, Harko, T., 2000, Mod. Phys. Lett.
A, 15, 2153

\bibitem{MaHaDo01} Mak, M. K., Dobson, P. N., Jr, Harko, T., 2001, Europhys. Lett., 55, 310

\bibitem{Bo04} B\"ohmer, C. G., 2004, Gen. Rel. Grav., 36, 1039;  A. Balaguera-Antolinez, C. G. B\"ohmer, M.
Nowakowski, 2004, preprint gr-qc/0409004

\bibitem{BoHa05}B\"ohmer, C. G., Harko, T., 2005, Phys. Rev., D71, 084026 

\bibitem{HaMa00} Harko, T., Mak, M. K., 2000, J. Math. Phys., 41, 4752

\bibitem{MaHa03} Harko, T., Mak, M. K., 2002, Annalen Phys., 11, 3; Mak, M. K., Harko, T., 2003, Proc. Roy. Soc. London,
A459, 393

\bibitem{FiSk95} Finch, M. R., Skea, J. E. F., 1995,  A Review of the Relativistic
Static Fluid Sphere, http://edradour.symbcomp.uerj.br/pubs.html

\bibitem{HeSa97} Herrera, L., Santos, N. O., 1997, Phys. Reports, 286, 53

\bibitem{DeLa98} Delgaty, M. S. R., Lake, K., 1998, Comput. Phys. Commun., 115, 395

\bibitem{LaLi} Landau, L. D., Lifshitz, E. M., 1995, The Classical Theory of
Fields, Butterworth Heinemann

\bibitem{HaHa94} Haggag, S., Hajj-Boutros, J., 1994, Class. Quantum
Grav., 11, L69


\bibitem{fern87} Fernley J. A., Jameson R. F., Sherrington M. R., Skillen I.,
1987, MNRAS, 225, 451

\bibitem{To39} Tolman, R. C., 1939, Phys. Rev., 55, 364

\bibitem{GuGu75} Gurses, M., Gursey, Y., 1975, Nuovo Cimento, 25B,
762

\end{thebibliography}
\end{document}